\documentclass[%
 aip,
 amsmath,amssymb,
 reprint,%
]{revtex4-1}

\usepackage{graphicx}
\usepackage{dcolumn}
\usepackage{bm}
\usepackage[utf8]{inputenc}
\usepackage[T1]{fontenc}
\usepackage{mathptmx}
\usepackage{etoolbox}
\usepackage{color}
\usepackage{longtable}
\usepackage{siunitx}
\usepackage{comment}
\usepackage{amsmath}

\makeatletter
\def\@email#1#2{%
 \endgroup
 \patchcmd{\titleblock@produce}
  {\frontmatter@RRAPformat}
  {\frontmatter@RRAPformat{\produce@RRAP{*#1\href{mailto:#2}{#2}}}\frontmatter@RRAPformat}
  {}{}
}%
\makeatother
\begin{document}

\preprint{APS/123-QED}

\makeatother

\title{Tunable Magnetic Skyrmions in Ferrimagnetic Mn$_4$N}
\author{Chung T. Ma}
    \email{ctm7sf@virginia.edu}
\affiliation{Department of Physics, University of Virginia, Charlottesville, Virginia, 22904, USA}
\author{Timothy Q. Hartnett}
\affiliation{Department of Materials Science and Engineering, University of Virginia, Charlottesville, Virginia, 22904, USA}
\author{Wei Zhou}
\affiliation{Department of Physics, University of Virginia, Charlottesville, Virginia, 22904, USA}
\author{Prasanna V. Balachandran}
\affiliation{Department of Materials Science and Engineering, University of Virginia, Charlottesville, Virginia, 22904, USA}
\affiliation{Department of Mechanical and Aerospace Engineering, University of Virginia, Charlottesville, Virginia, 22904, USA}
\author{S. Joseph Poon}
\affiliation{Department of Physics, University of Virginia, Charlottesville, Virginia, 22904, USA}
\date{\today}

\begin{abstract}

Thin films of ferrimagnetic Mn$_4$N are candidate materials to host magnetic skyrmions that have demonstrated thermal stability up to 450$^\circ$C.
However, there are no experimental reports observing skyrmions in this system. Here, we discuss the results of sputter grown 15-17~nm Mn$_4$N thin films on MgO substrate capped with Pt$_{1-x}$Cu$_{x}$ layers.
Vibrating sample magnetometry measurement of out-of-plane hysteresis loops confirmed that magnetic properties are insensitive to the cap layer composition.
Imaging based on magnetic force microscopy measurements observed 300 to 50~nm sized skyrmions, as the Cu concentration was increased from $x$ = 0 to 0.9. 
We performed density functional theory calculations and found that the interfacial Dzyaloshinskii-Moriya interactions (iDMI) follow a trend: Mn$_4$N/MgO(001) $<$ Cu/Mn$_4$N(001) $<$ Pt/Mn$_4$N(001).
We infer from these calculations that $x$ in Pt$_{1-x}$Cu$_{x}$ capping layer can serve as a robust tuning knob to tailor the iDMI and control the skyrmion size.
This work provides guidance to achieve smaller N\'{e}el-type skyrmions in Mn$_4$N thin films, which is an important step forward for building thermally stable skyrmionic devices.

\end{abstract}

\maketitle

As our society continues to increase its reliance on computing, efficient technologies are needed for processing and high-density data storage. Magnetic skyrmions are promising candidates to serve as information carriers in spintronics devices \cite{Tomasello2014racetrack,parkin2015racetrack,Kang2016Skyrmion_Electronics,Finocchio_2016_skm_funda,fert2017,kang2021_unconv}. Magnetic skyrmions are topological spin textures. They can be stabilized through the Dzyaloshinskii Moriya interaction (DMI) \cite{DZYALOSHINSKY1958,Moriya1960} or stray field \cite{Buttner2018}. In particular, DMI-skyrmions arise from broken inversion symmetry, which occurs in bulk materials with non-centrosymmetric crystal structure or thin films due to the interfacial asymmetry. Recent studies have reported Bloch skyrmions in bulk B20 crystals with intrinsic DMI at low temperature \cite{Jena2020b20,Zheng2018b20}. On the other hand, N\'{e}el skyrmions have been reported in heterostructures with heavy metal interfaces, such as Pt/CoFe/MgO \cite{woo2016exp}, Ir/Fe/Co/Pt \cite{Soumyanarayanan2017,Raju2019},  Pt/CoGd/Ta \cite{Caretta2018}, Pt/CoGd/Pt$_{1-x}$W$_x$ \cite{Quessab2020}, and (Pt/Co/Ru)$_x$ \cite{Legrand2020}. For reliable high-density data storage, small (near 10 nm) and stable skyrmions at room temperature are necessary. In ferromagnetic multilayers, due to their large saturation magnetization, which leads to the large stray field, only large skyrmions ($>$ 50 nm) have been observed at room temperature \cite{woo2016exp,Soumyanarayanan2017,Raju2019}. On the other hand, close to 10 nm skyrmions have been reported in amorphous ferrimagnetic rare-earth transition metal (e.g., GdCo) thin films \cite{Caretta2018}. However, their poor thermal stability presents a major challenge for device fabrication \cite{Hasegawa1974,Ueda2018}. 

One promising candidate with superior thermal stability compared to the amorphous rare-earth transition metals is rare-earth-free Mn$_4$N. There are some key similarities and differences between the rare-earth transition metals and Mn$_4$N.  
Both are ferrimagnetic metals, and there have been successful experimental demonstrations of thin film growth in both systems.
Both materials show perpendicular magnetic anisotropy (PMA) in the thin film geometry \cite{Shen2014,Yasutomi2014Mn,Kabara2015Mn4N,FOLEY2017Mnn4,Gushi2019Mn4N, Hirose2020Mn4N,Isogami2020Mn4n,Zhou2021_Mn4N, Quessab2020}.
However, unlike the amorphous structure in the rare-earth transition metals, Mn$_4$N is a crystalline compound that forms in the anti-perovskite crystal structure \cite{Zhou2021_Mn4N}.
Compared to the GdCo ferrimagnets, Mn$_4$N thin films offer key materials processing advantages for improving thermal stability due to the following two reasons: 
First, the Mn$_4$N films are deposited at 400-450$^\circ$C and no structural transitions or loss of PMA has been reported after annealing or cooling to room temperature. This means that Mn$_4$N films are robust to temperature changes during device processing. Second, PMA has been reported in films thicker than 10~nm that will enhance the lifetime of skyrmions \cite{Bessarab2018,Hoffmann2020}. Furthermore, recent experiment reports current-driven domain motion of up to 900 m/s for current density of 1.3 x $10^{12}$ J/$m^2$.  \cite{Gushi2019Mn4N} Such high mobility is crucial for high speed skyrmions motion.  
Although Mn$_4$N thin films appear to possess many favorable characteristics for hosting skyrmions, to date there is no experimental report in the open literature of observing skyrmions in Mn$_4$N thin films. In this paper, we report the experimental observation of skyrmions in Mn$_4$N thin films.
In a recent study, tuning of interfacial DMI (iDMI) was reported in Pt/CoGd/Pt$_{1-x}$W$_x$ through the dilution of Pt with W in the capping layer \cite{Quessab2020,golam2021gdco}. Based on the simulated skyrmion phase diagram, such tuning of iDMI was predicted to allow the control of skyrmion size for desirable applications \cite{Ma2019}.
In this work, we discuss the tuning of skyrmion size by diluting the Pt with Cu, which is a lighter element than W. Through imaging using magnetic force microscopy (MFM), skyrmions with sizes ranging from 300 to 50~nm have been observed in the MgO/Mn$_4$N/Pt$_{1-x}$Cu$_x$ heterostructures grown by magnetron sputtering. 
In addition, we also perform micromagnetic simulations and density functional theory (DFT) calculations to extract insights into the key characteristics that govern skyrmion size in the MgO/Mn$_4$N/Pt$_{1-x}$Cu$_x$ heterostructure. Similar to the Pt/CoGd/Pt$_{1-x}$W$_x$ work, our findings suggest that the Pt$_{1-x}$Cu$_x$ capping layer effectively reduces the iDMI as a function of $x$, which in turn leads to smaller skyrmions in the Cu-rich capping layer (eventually disappearing when $x$ = 0.95). This work shows a path for tailoring smaller skyrmions in Mn$_4$N thin films, whose properties compete with the well-studied amorphous GdCo and serves as a promising alternative for exploring skyrmion-based devices in thin film geometry with rare-earth-free elements.

We deposited 15$\pm$2~nm thick Mn$_4$N thin films on MgO(001) 5$\times$5$\times$0.5~mm substrate by reactive radio frequency (rf) sputtering at 450$^\circ$C. We also deposited 3~nm thick cap layers of Pt$_{1-x}$Cu$_{x}$ (where $x$ = 1, 0.5, 0.1, 0.05) on the sputter grown Mn$_4$N layer at room temperature by co-sputtering of Pt and Cu targets to tune the iDMI, and 3~nm Pt film on top as the capping layer to prevent oxidation. Details of the deposition process were reported earlier \cite{Zhou2021_Mn4N}. X-ray diffraction (XRD) was measured with Rigaku SmartLab. MFM images were taken with Bruker atomic force microscopy. Before MFM imaging, the Mn$_4$N samples were demagnetized by alternating and reducing the magnitude of the applied magnetic field. MFM images were analyzed by Gwyddion \cite{Necas2012Gwyddion} and magnetic moments were mapped out using the method described by Zhao \textit{et al} \cite{Zhao_2018MFM,Meng2019MFM}. We quantified the skyrmion size as the average diameter around a boundary where magnetization is zero \cite{romming2013exp,woo2016exp,Caretta2018}.

Spin-polarized electronic structure calculations were carried out in the DFT framework using the planewave pseudopotential code, \textsc{Quantum ESPRESSO} \cite{giannozzi2009quantum,giannozzi2017advanced,giannozzi2020quantum}. Core and valence electrons were treated using ultrasoft pseudopotentials \cite{Vanderbilt1990}. The exchange-correlation functionals were described using the Perdew-Burke-Ernzerhof parameterization of the generalized gradient approximation modified for solids \cite{PhysRevLett.100.136406}.
The plane wave cutoff energy was set to 60~Ry. We explored four $\Gamma$-centered Monkhorst-Pack $k$-point mesh \cite{MonkPack} to sample the Brillouin zone: 4$\times$4$\times$1, 10$\times$10$\times$1, 16$\times$16$\times$1 and 20$\times$20$\times$1. Our calculations revealed that we need at least a 16$\times$16$\times$1 $k$-mesh for convergence. Three separate interfaces were constructed to describe the MgO/Mn$_4$N/Pt$_{1-x}$Cu$_{x}$ heterostructure: Mn$_4$N/MgO(001), Pt/Mn$_4$N(001), and Cu/Mn$_4$N(001). The in-plane lattice parameter was fixed to the MgO substrate with a lattice parameter of 4.14~{\AA} for MgO. A vacuum of 22.5~{\AA} was added above the slab for both systems to limit interactions from the periodic image. In the Mn$_4$N/MgO(001) interface calculation, we assumed that the first Mn$_4$N atomic layer during thin film growth will contain only the Mn-atoms. In both the Pt/Mn$_4$N(001) and Cu/Mn$_4$N(001) calculations, the interface was also assumed to contain only the Mn-atoms. 

In both Mn$_4$N/MgO(001) and Pt/Mn$_4$N(001) calculations, the interfacial atoms, along with atoms in the layer directly above or beneath the interface, were allowed to relax in the $z$-direction to optimize the distance by minimizing the forces between the atoms. Since Mn$_4$N is a ferrimagnet in the inverse perovskite structure, spins of the two crystallographically unique Mn-sites that are away from the interface are antiferromagnetically coupled in our calculations. In the Cu/Mn$_4$N(001) calculation, we substituted all Pt atoms with Cu atoms, but maintained the same interatomic distances and structure, as a check on the impact of Cu-substitution. The atomic positions were relaxed until forces were less than 2 meV/{\AA} and the total energy converged to 10$^{-8}$~eV. 
The standard Kohn–Sham DFT with semi-local, gradient-corrected exchange-correlation functionals, generally underestimate the orbital magnetic moments due to the self-interaction error.\cite{LARSON20037, B_o_ski_2009, Flores_Livas_2019, ML_DFT_U} This error particularly leads to unphysical behavior in the partially-filled $d$-electron systems by incorrectly lowering the orbital energies and promoting spurious delocalization of the $d$-orbitals. One of the important consequences of this effect is the inaccurate description of the spin-orbit interaction, which will impact the iDMI calculations. We applied the localized Hubbard-$U$ correction, where we add a modest on-site Coulomb repulsion ($U_\textrm{eff}$=3~eV to Mn-$3d$ orbitals) to the potential acting on the Mn $3d$-electrons.\cite{PhysRevB.44.943, PhysRevB.52.R5467}.
For the total energy calculations that involved non-collinear magnetism with the spin-orbit coupling (SOC) term in the Hamiltonian, we set a convergence threshold of 10$^{-5}$~eV for the electronic steps. In the DFT-SOC-$U$ calculations, we used the fully-relativistic pseudopotentials only to describe the interfacial atoms directly above or beneath the interface.

\begin{figure}
\centering
  \includegraphics[width=\columnwidth]{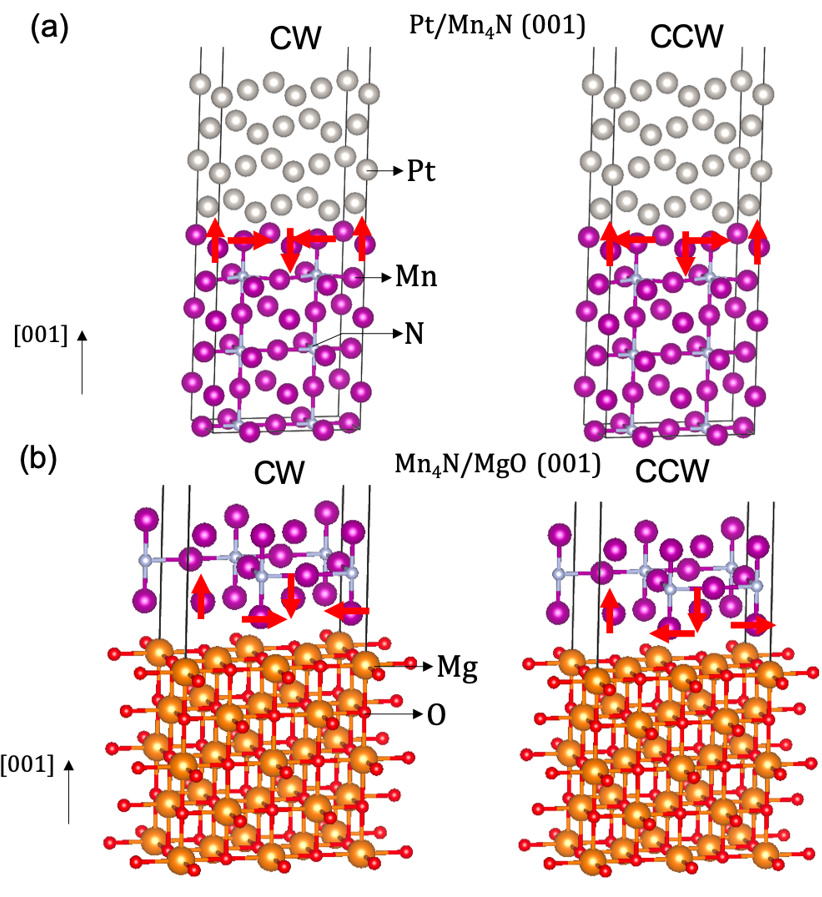}
  \caption{The atomic structures used in the DFT calculations. (a) Pt/Mn$_4$N(001) and (b) Mn$_4$N/MgO(001). The clockwise (CW) and counterclockwise (CCW) spin spiral arrangement on the interfacial Mn-atoms (purple color) are shown as red arrows. The spin spiral arrangement shown here corresponds to that of the $y$-orientation. In the Cu/Mn$_4$N(001) simulation, all Pt-atoms [shown in (a)] are substituted by the Cu-atoms.
  }
    \label{fig:dft_atoms}
\end{figure}
\begin{figure}
    \centering
    \includegraphics[width=\columnwidth]{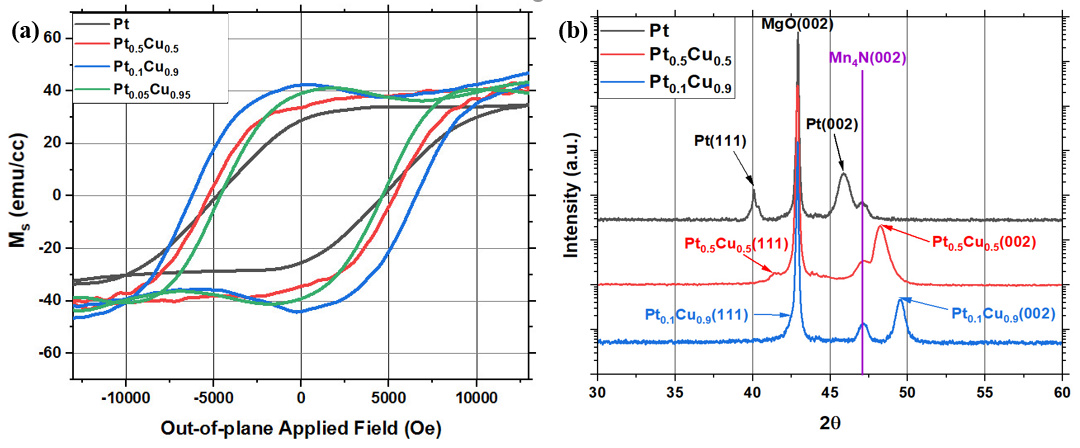}
     \caption {(a) Out-of-plane magnetic hysteresis loops of Mn$_4$N thin films measured at 300~K with four compositions of Pt$_{1-x}$Cu$_{x}$ cap layer. Results show similar magnetic properties in all four Mn$_4$N thin films, despite having different cap layer compositions. (b) X-ray diffraction (XRD) of MgO/Mn$_4$N(15 nm)/Pt$_{1-x}$Cu$_{x}$(30 nm)/Pt(3 nm). The y-axis of each XRD pattern is shifted to reveal differences in cap layer crystal structures.}
    \label{fig:Mn4N_MH}
\end{figure}

We followed the spin-spiral method developed by Yang \textit{et al}\cite{yang2015anatomy} to describe the non-collinear magnetism for the iDMI calculation. The spin spiral arrangements that was used in the DFT calculations for the $y$-orientation in the Pt/Mn$_4$N(001) and Mn$_4$N/MgO(001) interfaces are shown in Fig.~\ref{fig:dft_atoms}a and b, respectively. These calculations were generalized from the four-state energy mapping calculation of antisymmetric exchange interaction conducted by Xiang \textit{et al} \cite{xiang2011predicting, xiang2013energymapping}. The validity of this method has been explored by several authors in different magnetic materials classes, including interfaces and bulk materials which gives confidence in applying these approaches in the current work \cite{li2016mon, xu2018cri3, yuan2020magnets, akanda2020mnpt, sabani2020fourstate, golam2021gdco,yang2015anatomy}. In general, the energy due to DMI can be written as\cite{yang2015anatomy, li2016mon, akanda2020mnpt, sabani2020fourstate, golam2021gdco}, \begin{equation} E_{DMI} = \sum_{i,j} {\mathbf{d_{ij}}\cdot [\mathbf{S_i} \times \mathbf{S_j}]} + E_{other}\end{equation} where $\mathbf{d_{ij}}$ is the DMI vector, $\mathbf{S_i}$ and $\mathbf{S_j}$ are the unit vectors of a magnetic moment between the two nearest neighbor Mn-sites $i$ and $j$ in the Mn$_4$N structures (as shown in Fig.~\ref{fig:dft_atoms}), and $E_{other}$ is the energy due to all other interactions. Typically, $E_{other}=0$.\cite{yang2015anatomy, akanda2020mnpt} Based on this formalism, the total DMI strength ($D_{tot}^y$) concentrated in a single layer originating from the interaction of two interfacial Mn nearest neighbors can be written as\cite{yang2015anatomy}, \begin{equation} D_{tot}^y=\frac{\Delta{E}}{8\sqrt{2}} \label{eq:dm_vector} \end{equation} where $\Delta{E}$ = $E_{CW}-E_{CCW}$ is the DFT calculated total energy difference between the $CW$ and $CCW$ spin configurations, shown in Figure~\ref{fig:dft_atoms}. The denominator in Equation~\ref{eq:dm_vector} accounts for the contributions from the bond angle between adjacent Mn-atoms and the number of interacting Mn-atoms in the interfacial layer (in our simulation cell, this number is equal to 4). The superscript $y$ denote the spin spiral arrangement along the $y$-direction (as shown in Fig.~\ref{fig:dft_atoms}). Yang \textit{et al}\cite{yang2015anatomy} derived an equation connecting $D_{tot}^y$ from DFT to the micromagnetic DMI vector ($D_\mu^y$) that can be written as\cite{yang2015anatomy}, \begin{equation}  D_\mu^y = \frac{N_N \cdot D_{tot}^y}{N_L \cdot a^2} \label{eq:micro_dmi}\end{equation} where $N_N$ is the number of nearest neighbors to a Mn-atom, $N_L$ is the number of magnetic layers considered in the simulation cell with the non-collinear spin arrangement, and $a$ is the lattice parameter. We used $N_N=4$, $N_L=1$, and $a=4.14$~{\AA} in Equation~\ref{eq:micro_dmi}. We use the $D_\mu^y$ (given in Equation~\ref{eq:micro_dmi}) as a surrogate for the iDMI strength. 

We note that the purpose of these DFT calculations is not to accurately model the physics behind the MgO/Mn$_4$N/Pt$_{1-x}$Cu$_x$ interface. Our objective is to extract insights that will shed light on the expected iDMI trend at the three interfaces. 

Micromagnetic simulations were performed within the Object-Oriented Micromagnetic Framework (\textsc{OOMMF}) \cite{oommf}. Atomistic simulations were performed by an in-house package. The details of the simulations were reported in a previous publication \cite{Ma2019}, and supplementary materials. Although micromagnetic simulations do not take into account colinear spins in ferrimagnetic Mn$_4$N, it has been benchmarked with atomistic simulations for small skyrmions (< 40 nm) to verify its validity for a ferrimagnet.
In a previous publication of sub-20 nm Mn$_4$N films \cite{Zhou2021_Mn4N}, X-ray diffraction (XRD) showed only the MgO substrate peak and the Mn$_4$N (002) peak in the 2$\theta$-$\theta$ XRD profile. In the $\phi$ scan, four peaks at 90$^\circ$ interval confirms the epitaxial growth of Mn$_4$N (001)[100] on MgO(001)[100].
Fig.~\ref{fig:Mn4N_MH} (a) shows the out-of-plane hysteresis loops as measured by VSM. The data shows PMA in all four Mn$_4$N films with four cap layer compositions. All samples have a similarly saturated magnetization (M$_s$) of 50 emu/cc. Such small M$_s$ confirms the ferrimagnetic nature of Mn$_4$N. The remanent magnetization to M$_s$ ratio remains consistently larger than 0.7, insensitive to cap layer concentrations. This is a key result because it shows that the cap layer has little to no effect on the M$_s$, PMA, and remanent magnetization of the Mn$_4$N films. 
Fig.~\ref{fig:Mn4N_MH} (b) shows the XRD scans of MgO/Mn$_4$N(15 nm)/Pt$_{1-x}$Cu$_{x}$(30 nm)/Pt(3 nm). Thicker Pt$_{1-x}$Cu$_{x}$ cap layers are deposited to obtain an observable signal from the cap layers. Results show while the cap layers remain a face-centered cubic, lattice constant decrease as $x$ increases from 0 to 0.9.

\begin{figure}[!t]
\centering
    \includegraphics[width=\columnwidth]{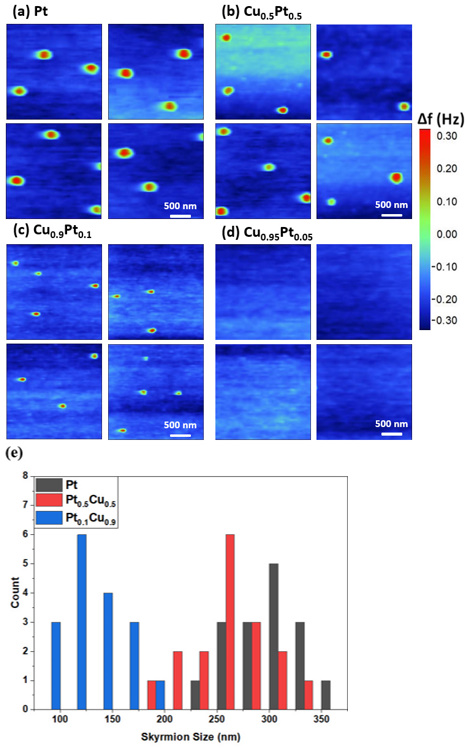}
     \caption{Room temperature magnetic force microscopy images of skyrmions in Mn$_4$N with various cap layers (a) Pt; (b) Pt$_{0.5}$Cu$_{0.5}$; (c) Pt$_{0.1}$Cu$_{0.9}$; (d) Pt$_{0.05}$Cu$_{0.95}$. In each figure, four areas of the same sample have been shown. Blue colors correspond to regions of down magnetization, which are the ferrimagnetic backgrounds. Red colors correspond to up magnetization, which is the core of a skyrmion.(e) A histogram of distribution of skyrmion size in Pt (Black), Pt$_{0.5}$Cu$_{0.5}$ (Red), and Pt$_{0.1}$Cu$_{0.9}$ (Blue), these distributions correspond to all areas imaged, in addition to the four areas shown.}
     \label{fig:Mn4N_skyrmion}
\end{figure}
We next focus on the MFM images that were taken for all four samples after undergoing the demagnetization process as stated in the Methods section.
Since the iDMI is related to the SOC \cite{Soumyanarayanan2016}, we expect the largest and smallest iDMI in the pure Pt and Pt$_{0.05}$Cu$_{0.95}$ cap layers, respectively.
From our previously published atomistic simulations work that explored the effect of iDMI on skyrmion sizes\cite{Ma2019}, we found that the skyrmions size is proportional to the iDMI strength. Therefore, we anticipate the MgO/Mn$_4$N/Pt and MgO/Mn$_4$N/Pt$_{0.05}$Cu$_{0.95}$ heterostructures to host the largest and smallest sized skyrmions, respectively.

Fig.~\ref{fig:Mn4N_skyrmion}(a)-(d) show the MFM images of skyrmions in Mn$4_N$ with different cap layers, and four different areas of a given sample are shown to demonstrate the distribution in skyrmion sizes. Fig.~\ref{fig:Mn4N_skyrmion} (e) shows the distribution of skyrmion size observed from different areas by MFM with Pt$_{1-x}$Cu$_{x}$ (where $x$ = 0.1, 0.9, 0.95) cap layer composition. A blue to red color code maps the frequency shift of the tip due to the interaction with magnetic moments within the sample. Blue color corresponds to negative frequency shift and negative magnetization, while red corresponds to positive frequency shift and positive magnetization. These regions were benchmarked using a demagnetized MgO/Mn$_4$N/Pt sample. The ferrimagnetic background of a magnetized Mn$_4$N film would give either a blue or red region, depending on magnetized direction. In this experiment, the Mn$_4$N films were first demagnetized then a down magnetic field is applied that gives the blue regions of ferrimagnetic backgrounds. Only spin textures that have a well-defined red core are considered skyrmions. For example, in the bottom left image of Fig.~\ref{fig:Mn4N_skyrmion}(c), the texture located on the top is not considered a skyrmion, because it lacks a well-defined core. A board distribution of skyrmion sizes are observed in these thin films, which are consistent with other experiments due to possible defects \cite{woo2016exp,Brandao2019mfm}.

In Fig.~\ref{fig:Mn4N_skyrmion} (a), for MgO/Mn$_4$N/Pt ($x$ = 0) thin film, the average skyrmion size is 300~nm. Our MFM work confirms that thin films of Mn$_4$N with a Pt heavy metal cap layer can host skyrmions, which is a promising result. To explore if we can further reduce the skyrmion size, we diluted the Pt heavy metal cap layer by co-depositing with Cu, which is a relatively lighter metal. 
Fig.~\ref{fig:Mn4N_skyrmion}(b) shows the MFM image of MgO/Mn$_4$N/Pt$_{0.5}$Cu$_{0.5}$. An average size of 250~nm skyrmion was observed with 50\% dilution of Cu. In Fig.~\ref{fig:Mn4N_skyrmion}(c), we show the MFM image for another heterostructure where we further increased the $x$ to 0.9 (Cu-rich). We find that the average skyrmion size reduced to 100~nm in MgO/Mn$_4$N/Pt$_{0.1}$Cu$_{0.9}$. Finally, we increased the $x$ to 0.95 resulting in a MgO/Mn$_4$N/Pt$_{0.05}$Cu$_{0.95}$ heterostructure with an almost pure Cu cap layer. In the MFM image shown in Fig.~\ref{fig:Mn4N_skyrmion}(d), only blue regions of ferrimagnetic background were observed, and no evidence for any skyrmions were found in MgO/Mn$_4$N/Pt$_{0.05}$Cu$_{0.95}$ heterostructure.
There are at least two plausible explanations that can help describe the disappearance of skyrmions in Fig.~\ref{fig:Mn4N_skyrmion}(d): (1) The iDMI from Pt$_{0.05}$Cu$_{0.95}$ is too weak to stabilize a skyrmion. (2) The limited spatial resolution of MFM fails to image smaller skyrmions ($<$ 50~nm). Further investigations with higher resolution imaging are needed to understand the complex physics at the Mn$_4$N/Pt$_{1-x}$Cu$_{x}$ interface. Nonetheless, these results show the unique potential of tuning skyrmions sizes in Mn$_4$N thin films by judiciously adjusting the cap layer composition. This outcome is expected to be critical for device applications, where smaller skyrmion sizes are desirable.
\begin{table}[htb]
\caption{\label{tab:idmi_DFT} DFT and DFT+$U$ (=3~eV) calculated $D_{tot}^y$ (in meV) and $D_\mu^y$ (in mJ/m$^2$) for the three different interfaces explored in this work.}

\begin{tabular}{l|l|l}
\hline
Interface & D$_{tot}^y$ (meV) & D$_{\mu}^y$ (mJ/m$^2$)\\
\hline
Mn$_4$N/MgO (001) & 0.290 & 1.082 \\
Mn$_4$N/MgO (001) (+$U$) & 0.273 & 1.017 \\
Pt/Mn$_4$N (001) & $-$3.20 & $-$11.957 \\
Pt/Mn$_4$N (001) (+$U$) & -1.86 & -6.969 \\
Cu/Mn$_4$N (001) & 2.602 & 9.710 \\
Cu/Mn$_4$N (001) (+$U$) & 0.706 & 2.633 \\

\hline
\end{tabular}
\end{table}

We now shift our attention to the DFT calculations. We obtained a significantly high D$_{tot}^y$ and $D_{\mu}^y$ in the Pt/Mn$_4$N(001) interface compared to the Mn$_4$N/MgO(001) interface. The data is given in Table~\ref{tab:idmi_DFT}. 
Interestingly, the Pt/Mn$_4$N(001) and Cu/Mn$_4$N(001) interfaces carry opposite $D_{tot}^y$ signs. Replacement of Pt with Cu in this interface resulted in a positive D$_{tot}^y$ value, but with a slight reduction in the magnitude. When the Hubbard-$U$ correction was added to electrons in Mn-$3d$ orbitals, we found a reduction in the calculated iDMI in all three interfaces. The most significant drop was in the Cu/Mn$_4$N (001) interface, where the iDMI dropped to 2.6 mJ/m$^2$. Compared to Pt, Cu is a light metal where we do not expect a large iDMI. This highlights the importance of the Hubbard-$U$ correction term in the iDMI calculations to address the self-interaction error.
The change of DMI sign has been observed before in Mn$_x$Fe$_{1-x}$Ge thin films, where it was shown to alter the helix chirality.\cite{PhysRevLett.110.207201, PhysRevMaterials.2.074404}
The consequence of the iDMI sign has been discussed previously in GdCo \cite{Ma2019}. The spins in a skyrmion will rotate in an opposite direction, but it is not detectable with MFM imaging.

Our DFT+$U$ calculated $|D_\mu^y|$ value for Pt/Mn$_4$N(001) interface is also of similar magnitude to other Mn compounds reported in the literature.
For example, Akanda \textit{et al} reported a $D_\mu$ value between $\sim$11 and 17~mJ/m$^2$ in the antiferromagnetic MnPt/W interfaces with a single magnetic layer configuration \cite{akanda2020mnpt}.
In another work, Yuan \textit{et al} reported a $D_\mu$ of $\sim$0, 9.1 and 6.7~mJ/m$^2$ in the monolayers of MnSSe, MnSTe and MnSeTe, respectively \cite{yuan2020magnets}. This was an intriguing result from the context of this work because it shows the importance of the neighboring coordinating atoms (S/Se/Te) in affecting the $D_\mu$ of the materials system.
In Mn$_4$N, we have lighter, non-magnetic neighboring N-atoms that are in coordination with the magnetic Mn-atoms.
More recently, Morshed \textit{et al} reported a $D_\mu$ between 0 and 4.5~mJ/m$^2$ in the  Pt(2)/GdCo(2)/Pt$_{1-x}$W$_x$(2) heterostructure, where (2) stands for the number of atomic layers in the simulation cell \cite{golam2021gdco}.
This comparison is also important because it shows that our DFT+$U$ calculated $D_\mu^y$ for Pt/Mn$_4$N(001) is larger than that of the Pt(2)/GdCo(2)/Pt$_{1-x}$W$_x$(2) heterostructure. This gives additional motivation for exploring the possibility of Mn$_4$N as a skyrmion host. 
The common thread linking the present work with that of Akanda \textit{et al}, Yuan \textit{et al}, and Morshed \textit{et al} is that all use the planewave pseudopotential-based DFT method within the four-state energy mapping analysis for simulating the spin spirals. 
There are also non-trivial differences between the specifics of the approaches adapted in the various papers and the present work. For instance, Morshed \textit{et al} and Yuan \textit{et al} include the Hubbard-$U$ term to describe the $3d$- and/or $4f$-electrons. However, Akanda \textit{et al} did not account for the Hubbard-$U$ term to describe the Mn-$3d$ electrons.
There are also other important differences in the choice of exchange-correlation functionals, supercell size, energy convergence threshold, inclusion of Hubbard corrections (Hubbard-$U$), and $k$-point mesh that may affect the calculated $D_\mu$ data.
It is also important to recognize that all DFT calculations are performed at 0~K. Systematic experimental measurements on various interfaces have revealed a strong temperature dependence of iDMI.\cite{Schlotter_DMI_T_2018, zhou2020temperature} Typically, iDMI increases with decreasing temperature. Therefore, we anticipate the room temperature iDMI values to be smaller in magnitude compared to the data given in Table~\ref{tab:idmi_DFT}.
\begin{figure}
    \centering
    \includegraphics[width=\columnwidth]{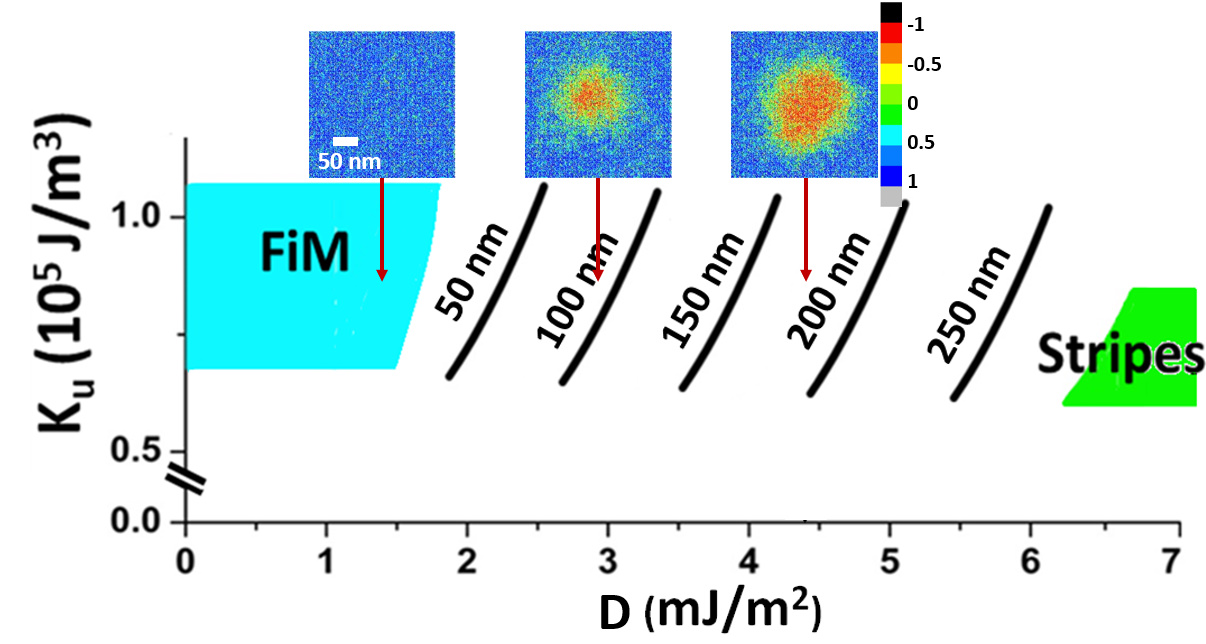}
     \caption{Simulated skyrmion DMI-K$_u$ phase diagram of 15-nm Mn$_4$N at 300 K. Dilution Pt cap layer by Cu reduces iDMI, which leads to smaller skyrmions. Inserted figures above show the mapping of out-of-plane magnetization from simulations. Red region corresponds to down magnetic moments while blue region corresponds to up magnetic moments. For an iDMI in the range from 3.5 to 4 mJ/m$^2$, a close to 200 nm skyrmion is stabilized at 300~K. If the iDMI is reduced to 3.0 mJ/m$^2$, skyrmion size is reduced to 100 nm. If the iDMI is further reduced to 1.5 mJ/m$^2$, skyrmions are unstable at room temperature and collapse into ferrimagnetic state.} 
    \label{fig:Mn4N_D_K}
\end{figure}

In Fig.~\ref{fig:Mn4N_D_K}, we show a simulated D$_\mu$-anisotropy (K$_u$) skyrmion phase diagram for Mn$_4$N thin films from the micromagnetics simulations.
We systematically varied the iDMI from 0 to 7~mJ/m$^2$ to map the phase diagram.
In the D-K diagram, ferrimagnetic states are found for iDMI below 1.5 mJ/m$^2$. As iDMI increases from $\sim$ 1.5 mJ/m$^2$ to $\sim$ 6 mJ/m$^2$, skyrmion sizes increase from 50 nm to 250 nm. Above 6 mJ/m$^2$, although stripes are observed, larger skyrmions ($>$ 250 nm) can still exist, as the in-plane simulated space of 300 nm $\times$ 300 nm limits the simulated skyrmion size.
In addition, the micromagnetics simulations predict that as the D$_\mu$ decreases the skyrmion size should also decrease from 200 nm to as small as 50~nm (under the constraint that all other parameters are held to a constant value). 
These results support our expectation based on the Pt/CoGd/Pt$_{1-x}$W$_x$ work \cite{Quessab2020} that dilution of cap layer using a lighter metal, such as Cu, can be effective in stabilizing smaller skyrmions by reducing the iDMI strength. 
In addition to the cap layer, the Mn$_4$N materials family offers immense scope for further tuning the skyrmion sizes via alloying strategies.

Finally, we address some of the plausible reasons for the observed discrepancies between the predicted vs experimentally measured skyrmion size: (1) The iDMI parameter from DFT was calculated at 0~K, whereas the atomistic simulations and experimental measurements were performed at 300~K. Zhou \emph{et al} and Schlotter \textit{et al}  have shown that iDMI is sensitive to temperature by experimental measurements.\cite{Schlotter_DMI_T_2018,zhou2020temperature} (2) In our DFT calculations, we assumed an ideal interface of Mn-atoms. However, the actual atomic structure near the interface is not known. (3) Uncertainty in the four-state energy mapping analysis is not known and how it propagates to the micromagnetics simulations can impact the predicted skyrmion size. (4) Defects in the thin film sample due to experiments and materials processing can also lead to variations in iDMI, and thus the skyrmion sizes \cite{Caretta2018}.
Our study motivates the need for more experimental and theoretical studies to fully understand the intriguing behavior of rare-earth-free Mn$_4$N thin films for skyrmion-based applications.

In summary, thermally stable Mn$_4$N thin films have been investigated as a potential material for skyrmion-based spintronics. Magnetic skyrmions in Mn$_4$N films stabilized by varying iDMI are imaged by magnetic force microscopy. Skyrmion sizes can be tuned from 300 nm to 80 nm in MgO/Mn$_4$N/Pt$_{1-x}$Cu$_{x}$ by increasing Cu concentration ($x$) from 0 to 0.95. DFT calculations are employed to study the iDMI in three distinct interfaces. Micromagnetic and atomistic simulations informed by the experimental and DFT calculated parameters shed light on the plausible explanation behind tuning of the skyrmion size in Mn$_4$N thin films, and suggest a potential for further reduction in size to $<$ 50~nm with judicious interfacial engineering. These results provide a promising outlook for tailoring skyrmions in Mn$_4$N-based thin films, with implications in enabling future spintronics technologies.

This work was supported by the DARPA Topological Excitations in Electronics (TEE) program (grant D18AP00009). The content of the information does not necessarily reflect the position or the policy of the Government, and no official endorsement should be inferred. Approved for public release; distribution is unlimited.

The data that support the findings of this study are available from the corresponding author upon reasonable request.

Accepted for publication in Applied Physics Letters.

 \bibliography{ref}

\section{Supplementary materials}
\subsection{Methods}
\subsubsection{Experiments}
\begin{figure}
    \centering
    \includegraphics[width=\columnwidth]{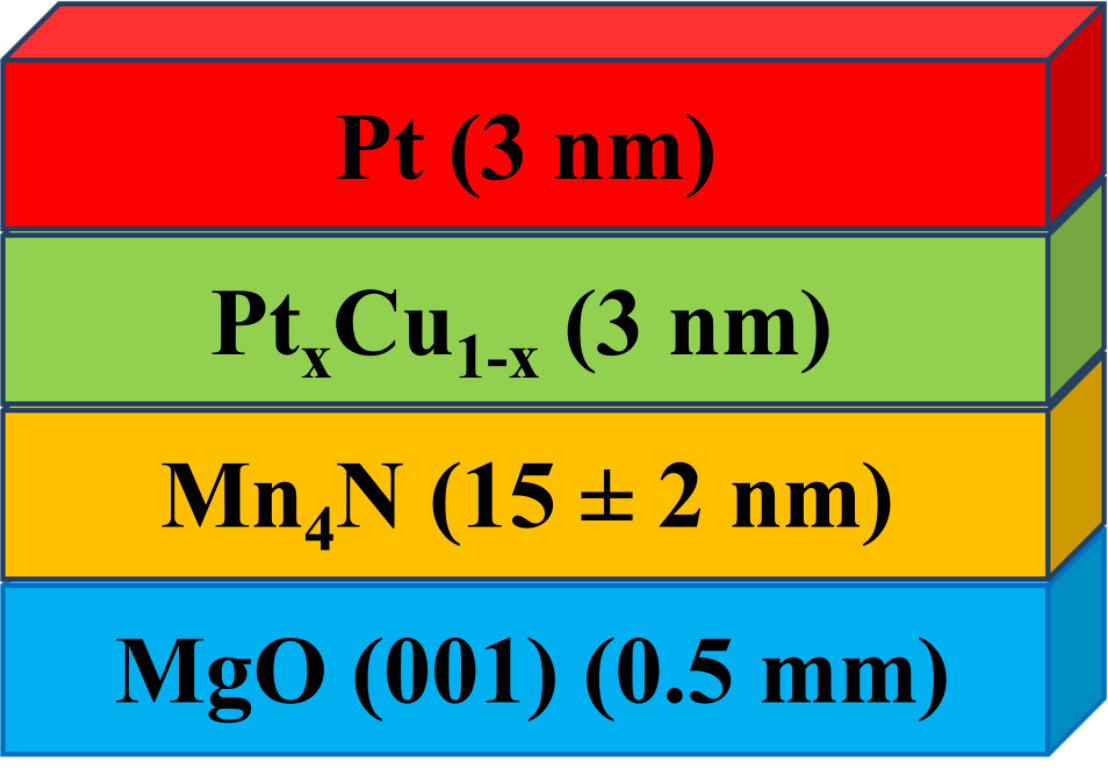}
     \caption{Schematic diagram of Mn$_4$N/Pt$_{1-x}$Cu$_x$ thin films
deposited on MgO(001) substrates.} 
    \label{fig:stack}
\end{figure}

For deposition of Mn$_4$N, we used a base pressure of 7 x 10$^{-8}$ Torr and deposited from a single Mn target. Flow rates of Ar:N$_2$ ratio were maintained at 93:7. MgO substrates were wet-cleaned and heat-treated before loading into the vacuum chamber. Details of the pre-sputtering cleaning process and sputtering process were reported earlier\cite{Zhou2021_Mn4N}. Fig. \ref{fig:stack} shows an illustration of these heterostructures. The film thickness and epitaxial growth of Mn$_4$N were verified by X-ray reflectometry and diffraction with Rigaku SmartLab. Magnetic properties were measured with VersaLab vibrating sample magnetometry (VSM).

\subsubsection{Micromagnetic Simulations}

Both micromagnetic and atomistic simulations were carried out at room temperature using Landau-Lifshitz-Gilbert (LLG) equation, also discussed in earlier work. Total simulated space is 300 nm x 300 nm x 15 nm in micromagnetic simulations, and each cell size is 5 nm x 5 nm x 5 nm. For these simulations, the exchange stiffness constant (A) is 1.5 x 10$^-11$ J/m, estimated from Curie temperature of 745 K\cite{Gushi2019Mn4N}. Saturation magnetization Ms is 50 kA/m and anisotropy K$_u$ is between 0.7 and 1.1 x 10$^5$J/m$^3$, based on the previous measurements \cite{Zhou2021_Mn4N}.

\subsection{Results}
\subsubsection{Additional Magnetic Properties of Mn$_4$N}
\begin{figure}
    \centering
    \includegraphics[width=\columnwidth]{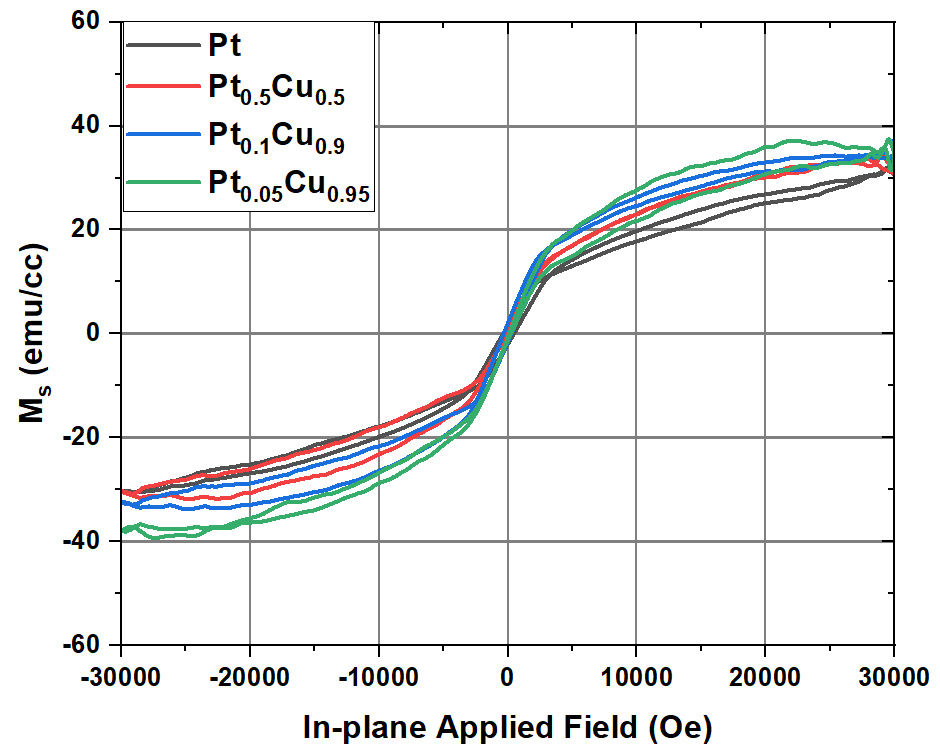}
     \caption{Room temperature in-plane hysteresis loop of Mn$_4$N with Pt$_{1-x}$Cu$_x$ cap layer.} 
    \label{fig:inplaneMH}
\end{figure}
Since anisotropy (K$_u$) plays a role in skyrmion formation, any changes in K$_u$ will also affect the skyrmion size. To show that changes in skyrmion size observed are not due to K$_u$ differences, in-plane hysteresis loop measurements of Mn$_4$N with Pt$_{1-x}$Cu$_x$ cap layer are measured and shown in Fig. \ref{fig:inplaneMH}. Results show a similar anisotropy field (> 3 T) in Mn4N. This verifies that the anisotropy (K$_u$) of Mn$_4$N is independence of cap layers. Thus, the change in skyrmion size is not due to the change in K$_u$.

\end{document}